# Reward based optimization of resonance-enhanced piezoresponse spectroscopy

Yu Liu[1*], Boris Slautin[2], Jason Bemis[3], Roger Proksch[3], Rohit Pant[4], Ichiro Takeuchi[4], Stanislav Udovenko[5], Susan Trolier-McKinstry[5], and Sergei V. Kalinin[1,6*]

[1] Department of Materials Science and Engineering, University of Tennessee, Knoxville, Tennessee 37996, USA

[2] Institute for Materials Science and Center for Nanointegration Duisburg-Essen (CENIDE), University of Duisburg-Essen, Essen, 45141, Germany

[3] Oxford Instruments Asylum Research, Santa Barbara, California, 93117, USA

[4] Department of Materials Science and Engineering, University of Maryland, College Park, Maryland 20742, USA

[5] Materials Science and Engineering Department, Materials Research Institute, the Pennsylvania State University, University Park, Pennsylvania, 16802, USA

[6] Physical Sciences Division, Pacific Northwest National Laboratory, Richland, Washington, 99354, USA

* Corresponding author: yliu206@utk.edu, sergei2@utk.edu

**Abstract**

Dynamic spectroscopies in Scanning Probe Microscopy (SPM) are critical for probing material properties, such as force interactions, mechanical properties, polarization switching, and electrochemical reactions and ionic dynamics. However, the practical implementation of these measurements is constrained by the need to balance imaging time and data quality. Signal to noise requirements favor long acquisition times and high frequencies to improve signal fidelity. However, these are limited on the low end by contact resonant frequency and photodiode sensitivity, and on the high end by the time needed to acquire high-resolution spectra, or the propensity for samples degradation under high field excitation over long times. The interdependence of key parameters such as instrument settings, acquisition times, and sampling rates makes manual tuning labor-intensive and highly dependent on user expertise, often yielding operator-dependent results. These limitations are prominent in techniques like Dual Amplitude

Resonance Tracking (DART) in Piezoresponse Force Microscopy (PFM) that utilize multiple concurrent feedback loops for topography and resonance frequency tracking. Here, a reward-driven workflow is proposed that automates the tuning process, adapting experimental conditions in real time to optimize data quality. This approach significantly reduces the complexity and time required for manual adjustments and can be extended to other SPM spectroscopic methods, enhancing overall efficiency and reproducibility.

**Introduction**

Dynamic spectroscopies play a crucial role in Atomic Force Microscopy (AFM), encompassing a variety of techniques such as amplitude and phase detection during force-distance measurements [1] and electromechanical spectroscopies of ferroelectric and ionic materials. These approaches are valuable in specialized AFM modes like Piezoresponse Force Microscopy (PFM) [2] and Electrochemical Strain Microscopy (ESM) [3, 4], where a broad range of spectroscopic measurements can be employed. These techniques allow researchers to probe intricate material behaviors, such as polarization dynamics, reversible and irreversible electrochemical reactions, and more. Such capabilities are essential for studying the functional properties of ferroelectrics, piezoelectrics, and electrochemical materials at the nanoscale.

In PFM and ESM spectroscopies, application of a periodic bias to the probe results in surface deformations, that are further explored as a function of time and tip bias, often following complex biasing waveforms designed to separate specific materials responses [5, 6]. When implemented in the classical beam-deflection set-up, the small bias-induced strains necessitates the use of resonance amplification methods [7, 8]. The resonance frequency of the SPM tip in contact with ferroelectric or electrochemically active material can in turn be bias dependent. The latter provides information on phenomena such as polarization rotation, in multiaxial ferroelectrics [9, 10], transition between ergodic and non-ergodic phases in relaxors [11-14], composition dependence of elastic moduli in electrochemical systems [15-20], or surface deformation due to the local irreversible electrochemical reaction [21-23].

It is important to note the role of non-linear interactions. For ferroelectric materials, of special interest is the evolution of electromechanical response with the driving voltage, containing information on the domain wall dynamics and intrinsic ferroelectric nonlinearities. Several studies reporting these effects in SPM measurements have been reported [24, 25]. In bulk materials, these effects are often coupled with the elastic non-linearities, manifesting as deviations from simple harmonic oscillator type behaviors. In SPM, the studies of such behaviors have been reported by Vasudevan [26]. However, the deviation from linear model were found to be small; furthermore, non-linearities in frequency domain can also be associated with the non-linear mechanical behaviors of the tip-surface junction, esp. for small indentation forces and large driving amplitudes. Here, we assume the intrinsic mechanical non-linearities are small.

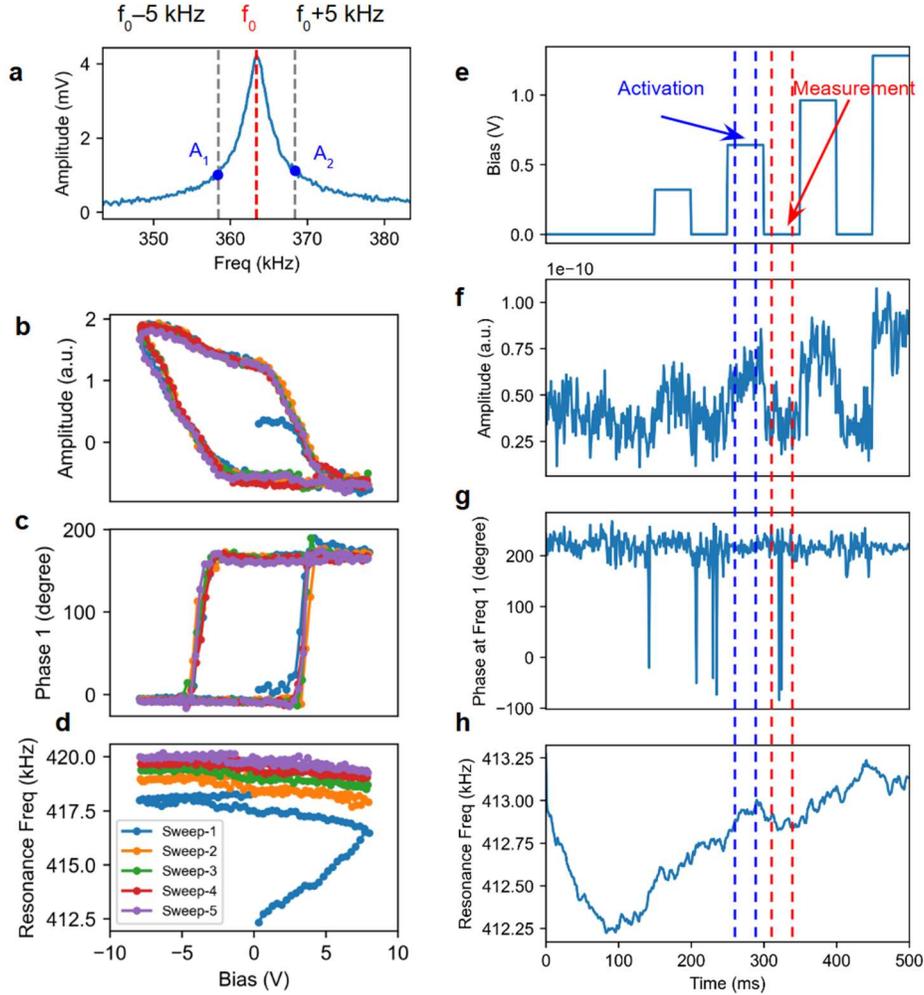

**Figure 1. Overview of the DART measurement. a,** tuning curve of a probe in the DART mode. Two amplitudes, $A_1$ and $A_2$, are measured at two frequencies located on different sides of the resonance frequency. By maintaining a constant difference between the two amplitudes ($A_1 - A_2$), DART tracks the resonance frequency precisely and removes artifacts induced by drift of the resonance frequency. **b-d,** typical DART spectra of the piezo response **(b)** amplitude, **(c)** phase and **(d)** resonance frequency. **e-h,** time sequence of the DART measurement. Each measurement cycle consists of an activation period at a specific bias followed by a measurement period at zero bias. The **(f)** amplitude, **(g)** phase and **(h)** frequency are measured by averaging the corresponding signals in the measurement period.

The two primary paradigms for the resonance enhancement in PFM are band excitation (BE) [8] and dual amplitude resonance tracing (DART) [21]. In the case of BE, the probe is excited by the signal having tailored, usually uniform, spectral density in the frequency band encompassing the cantilever resonance. The Fourier transform of the response signal yields the segment of the amplitude- and phase-frequency curve. The excitation band is chosen to be sufficiently broad to accommodate possible changes in the resonance frequency with position or bias [27]. The important advantage of band excitation is that the full amplitude frequency curve can be measured, further providing insight into the non-linear frequency responses [26].

Alternatively, the band of excited frequencies can be actively varied to follow a resonance that is moving. DART also uses active detection, where the probe is excited simultaneously at two frequencies on either side of the resonance (shown in Figure 1a). The corresponding amplitudes are determined using standard lock-in detection, and the difference of the amplitudes is used as feedback to track the resonance between excitation frequencies.

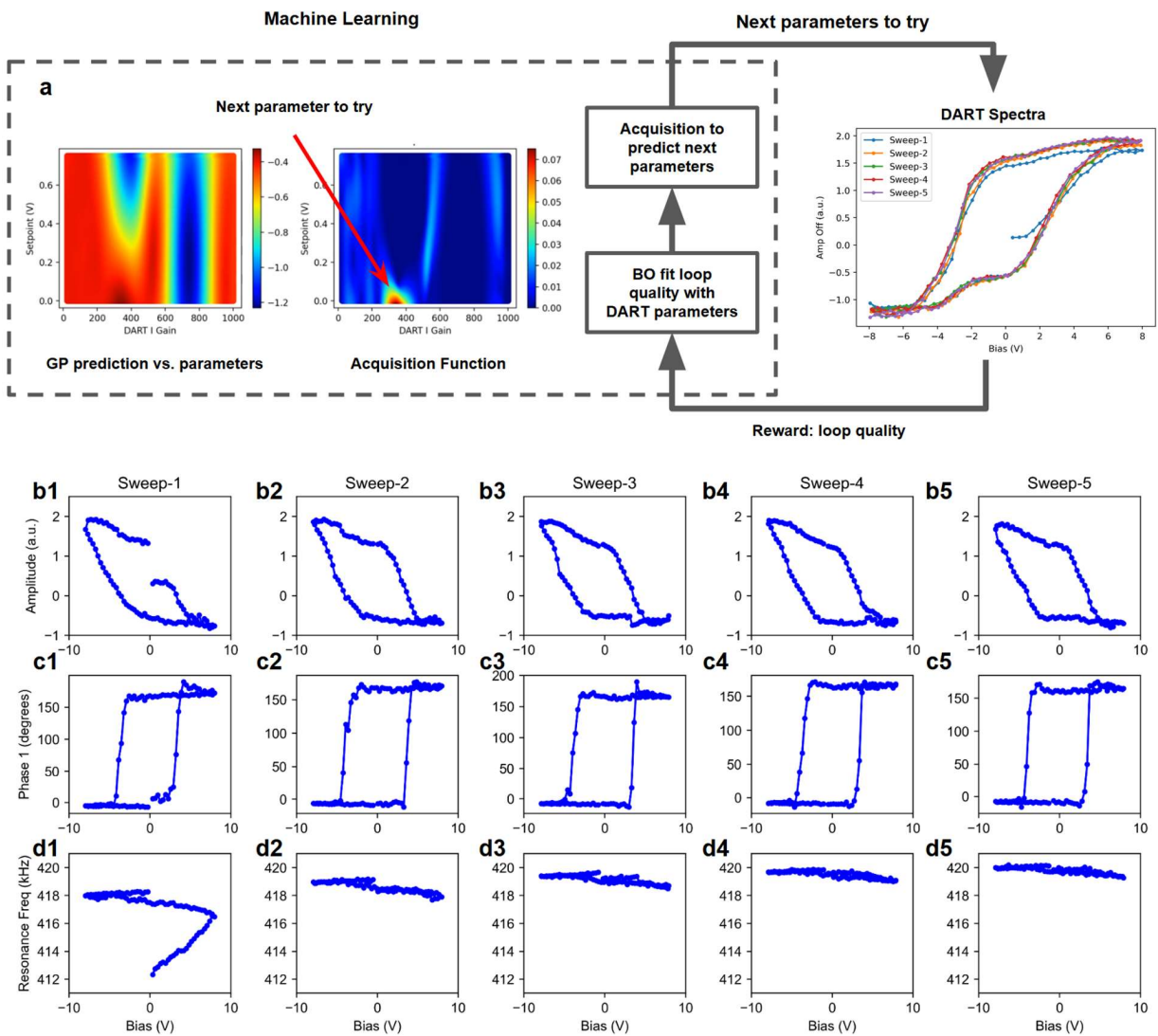

**Figure 2. Workflow of autonomous optimization of DART parameters. a,** in each iteration of the autonomous exploration, a reward function quantifying the loop quality is extracted from the DART spectra. In the Bayesian optimization (BO) workflow, a Gaussian process (GP) model is used to fit the reward function with the DART parameters, from which an acquisition function is computed. The next DART parameters to explore are determined by maximizing the acquisition function. This exploration process is repeated until the threshold precision or the maximum number of experiment steps is reached. Ferroelectric response **(b1-b5)** amplitude curves, **(c1-c5)** phase curves, and **(d1-d5)** resonance frequency curves of the five cycles measured at the same location with DART. The reward function is computed based on the consistency of amplitude curves from the second sweep to avoid the large resonance frequency drift in the first sweep.

Both for BE and DART spectroscopies, the optimization of the experimental parameters can be a significant challenge. From the operator viewpoint, the desired outcome is data on a dense spatial grid with sufficiently high signal to noise ratio and small instrument distortions. Practically, instrument stabilities, tip damage, sample damage and the total experiment time limit data acquisition times to several hours per scan. Combined with the need to sample the spatial domain at (50-100) [2] pixels, this corresponds to ~seconds per spectroscopic waveform. At the same time, from an instrument perspective the lowest accessible times in the system are those of the response time of resonance tracking. Correspondingly, the probing waveforms including multiple hysteresis loops, first order reversal curves, bias-dependent relaxations, etc., have to be accommodated within this frequency band.

For DART, these considerations require careful tuning of the parameters including sampling in the bias space, gains for the DART feedback and topographic feedback, and other control parameters as shown in Figure 1. Often the effects of these controls are coupled in difficult to predict ways. For example, electrostatic forces can result in the cantilever deflection and apparent height shifts, that can be compensated by the piezoelectric based motion if active topographic feedback is engaged during the measurements. As a result, DART optimization is time consuming and strongly operator dependent. While single point measurements can be made reproducible through selection of sufficiently large acquisition times, spectroscopic mapping often is not reproducible due to the multiple transient effects. These behaviors are illustrated in Figure 1 e-f, showing the transients associated with the detected amplitude and frequency shift signals during the applications of the spectroscopic waveform and associated noise levels. Once averaged, these transients manifest as the irreproducibility and drift in the hysteresis loop change and its dependence on imaging parameters. Hence, the goal of optimization workflows is to minimize the loops distortions while minimizing the acquisition times.

Here, ML-driven optimization of DART PFM is demonstrated using a reward-driven approach. In this, the criteria used by human operators for optimization are elucidated and converted into a quantifiable form. A reward function is then defined to quantify the quality of DART spectroscopy based on the consistency of the piezoresponse amplitude hysteresis loops, which is computed based on the normalized standard deviation across different sweeps in the amplitude curves. With this reward function defined, the problem is reduced to that of the optimization of reward function in the instrument control parameter space.

To implement reward driven optimization, the first step is to define the instrument parameter spaces. These are generally provided by the manufacturer and include probe related parameters like excitation bias, setpoint deflection, and I (integral) gain for the z-feedback loop. In addition, there are also parameters for the resonance frequency tracking and DART spectroscopy, which include DART I gain, DART frequency width, sampling rate and low-pass filter frequency, spectroscopy time, bias sweeping range and resolution, etc. Among these parameters, the deflection setpoint determines how hard the cantilever is pressed down, which has different optimal values for different sample surface quality. The DART I gain parameter controls how fast the DART responds to a change of resonance frequency, which is crucial for a successful DART measurement. Thus, the deflection setpoint and DART I gain were chosen as the parameters to optimize. The autonomous optimization workflow can be expanded with more parameters.

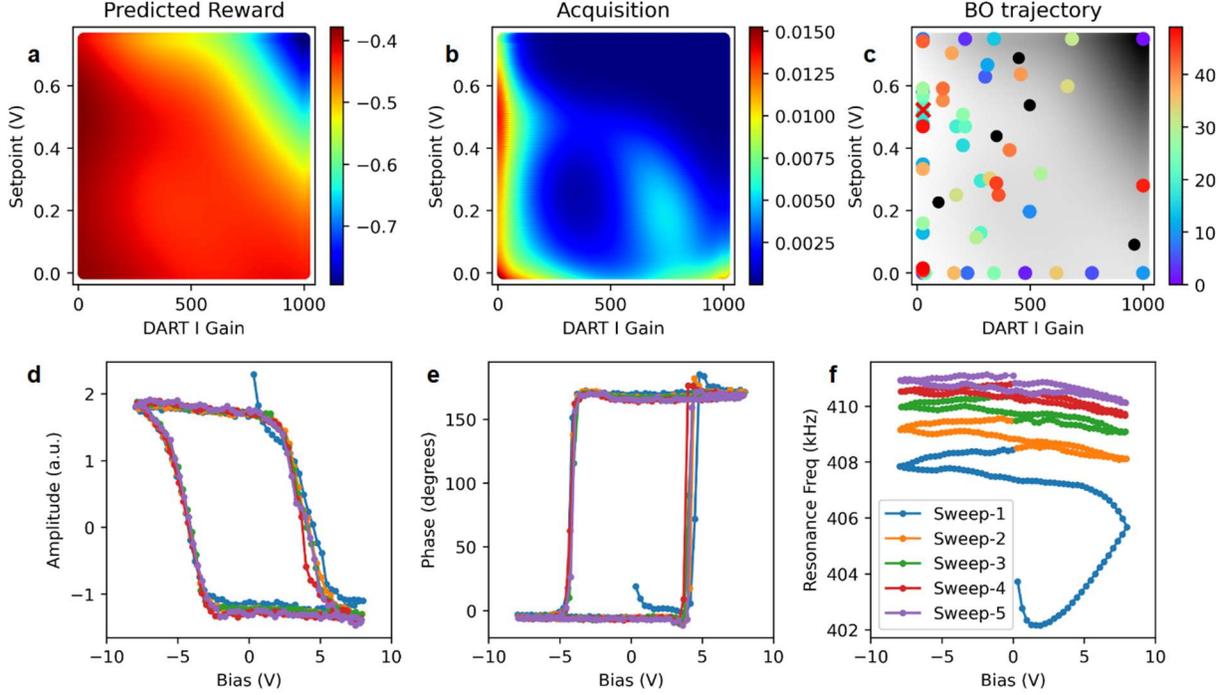

**Figure 3. Optimization process of DART parameters on a ~150 nm thick Pb$_{0.995}$(Zr$_{0.45}$Ti$_{0.55}$)$_{0.99}$Nb$_{0.01}$O$_3$ (PZT) film. a,** predicted reward function after 5 seeding points and 30 exploration steps. A higher reward function means the loop measurements are more consistent across different sweeps. **b,** based on the GP model, an acquisition function of expected improvement (EI) is calculated to decide the next parameters to try, which are the parameters corresponding to the maximum BO acquisition function value. **c,** the trajectory of the exploration is plotted on top of the predicted reward. The underlying white-gray color scale shows the predicted reward same as a. **d-f,** DART spectra of piezo response **(d)** amplitude, **(e)** phase, and **(f)** resonance frequency measured with the optimized DART parameters of setpoint=0.523 V and DART I Gain=25 given by the GP model in **c**.

The reward function is further defined to quantify the quality of the DART measurements. Here we note that the definition of the reward function can be application specific. First, the characteristics of good DART measurements are examined. For a high-quality piezoelectric response hysteresis loop, the amplitude curves should be consistent across different sweeps, indicating that there is minimal electrostatic effect induced by the measuring bias and the spectrum reflects the electromechanical properties of the materials. Therefore, the consistency of the amplitude channels across different sweeps is defined as the reward function:

$$\text{Rewards} = \sum_{i=2}^{n}(|A_i - A_{ave}|)/|A_{0i}|$$

where $A_i$ is the ferroelectric response amplitude curve of the $i$-th cycle, $A_{ave}$ is the averaged ferroelectric response amplitude curve of all the cycles except for the first one. $|A_{0i}|$ is the maximum amplitude of the $i$-th cycle and serves as a normalization factor. Notice that the rewards are computed based on the amplitude curves from the second cycles because usually the first cycle suffers from a large drift of the resonance frequency and the induced large deviation of the amplitude, as shown in Figure 2.

With the reward function defined, the DART on a real SPM instrument was optimized. The Automated Experiments on SPM (AESPM) package was used to automatically tune DART parameters and conduct DART measurements with Python codes [28]. The optimization is performed with the BOTorch implementation of Bayesian Optimization (BO) [29]. Bayesian optimization uses a Gaussian process (GP), a probabilistic model, to build a surrogate model of the reward function. This surrogate model predicts the function's behavior and provides uncertainty estimates of its predictions. The optimization process then uses this model to decide where to sample next in order to balance exploration (sampling where uncertainty is high) and exploitation (sampling where the predicted value is promising).

In the autonomous optimization of DART, the workflow starts with acquiring 5 seeding DART spectra at randomly selected parameters. In the optimization process, each iteration starts with fitting the reward function extracted from the acquired DART spectra with the DART parameters. Then an acquisition function of expected improvement (EI) will be computed based on the prediction and uncertainty of the fitted GP model. The next parameter to try is determined by finding the DART parameters that give the maximum BO acquisition function value and will be used in the next iteration. The optimization process finishes when the threshold reward function or the maximum number of exploration steps of 30 is reached.

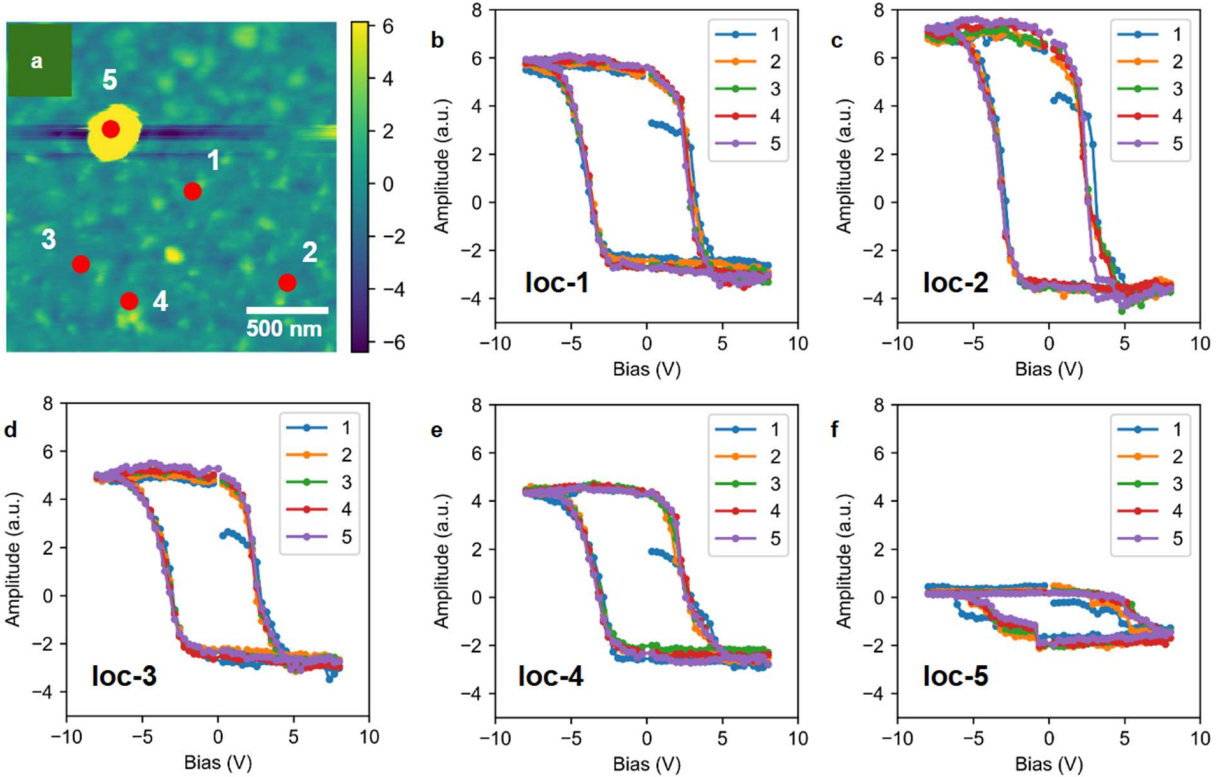

**Figure 4. Consistency of DART data on $Sm_{0.1}Bi_{0.9}FeO_3$ (SmBFO). a,** topography map taken with DART mode in contact shows the structural features around the location of optimization. Autonomous optimization was conducted at location-1. **b-f,** piezoelectric response amplitude hysteresis loop measured with the autonomously optimized DART parameters at locations 1-5, respectively. The optimized DART parameters give DART spectra with similar quality from location 1 to 4, where the sample surface quality is similar. At location 5, the DART spectra show less optimal consistency because there is a higher boulder.

Figure 3 shows an example optimization process conducted on PZT. After 5 seeding steps and 30 optimization steps, the GP model predicts the distribution of loop quality across the parameter space as shown in Figure 3a. Based on this prediction and the uncertainty of the GP model, it determines the optimal DART parameters that give the DART measurements with highest quality in Figure 3c. Indeed, the DART spectra measured with the optimal parameters show high-quality and consistent results in Figure 3d-f.

To verify the consistency of the optimization process, the DART data were measured with the optimized parameters at different locations. The DART parameters were optimized on SmBFO, at the center of a 2 μm area shown in Figure 4a. Three locations (location 2-4 in Figure 4a) with similar sample quality as location 1 were then randomly selected within the same 2 μm map and an additional location was chosen with a different topography (location 5 in Figure 4a). The measured DART spectra shows that provided the sample quality is similar to the location of optimization, the optimized parameters will give hysteresis loop spectra with similar quality. However, once the sample surface quality has changed, the DART parameters have to be re-optimized.

It is further noted that the DART optimization problem can be formulated differently depending on the experimental goal. Here, we develop a sample-agnostic version by optimizing the instrument parameters for a given time per loop. This assumes that the total measurement time and optimal spatial sampling are determined by the operator, and the experiment goal is full knowledge of the loop. Alternative settings can be optimization for the best value, balancing the level of details in a single loop against the number of spatial points. In this case, the time per loop becomes the optimization parameter; however, optimization requires assigning a certain value to the level of spectroscopic and spatial details. Correspondingly, optimization can be defined only at the level of multiple scan lines, i.e. large feedback times. Alternatively, the spectroscopic waveform can be optimized for determination of specific parameters such as nucleation bias. This approach requires reconfigurable probing waveforms and ML algorithms that incorporate prior knowledge in measurements (since for open priors uniform sampling of voltage space is the optimal strategy). Finally, experiments can be configured for mapping [30, 31] or discovery [32, 33] of specific microstructural elements, necessitating more complex imaging workflows.

In summary, a reward-driven workflow was implemented to automatically optimize the DART parameters to achieve high quality spectroscopic measurements. The reward function was defined based on the consistency of the piezoresponse amplitude curves. Because this reward function captures the characteristics of good DART measurements and is independent of the choice of the probes and samples, the autonomous workflow gives consistently good results on PZT and different locations on a SmBFO combinatorial library sample. The workflow makes it possible to apply DART-SPM in a standardized way in studies on ferroelectrics or electrochemical studies. This work also paves the way for building more complicated imaging workflows that include DART optimization as a component.

## Methods

Pb$_{0.995}$(Zr$_{0.45}$Ti$_{0.55}$)$_{0.99}$Nb$_{0.01}$O$_3$ films were grown by pulsed laser deposition using a KrF excimer laser from a ceramic target onto a SrRuO$_3$-electroded (001) SrTiO$_3$ single crystal. The SrRuO$_3$ film was grown from a target from Kojundo Chemical Lab. Co. Ltd., using a laser energy density of 1.5 J/cm$^2$, a substrate temperature of 660°C, an oxygen pressure of 120 mTorr, a target-to-substrate distance of 6.7 mm, and a frequency of 5 Hz. The SrRuO$_3$ film thickness was around 50 nm. The PZT film was grown from a target with 20% excess PbO to compensate for lead loss during growth, using a laser energy density of 1.5 J/cm$^2$, a substrate temperature of 630°C, an oxygen pressure of 120 mTorr, a target-to-substrate distance of 6.2 mm, and a frequency of 5 Hz. The PZT film thickness was around 147 nm.

The Sm-doped BiFeO$_3$ (SmBFO) thin film used in this paper was grown on a SrTiO$_3$ substrate, with 10% Bi replaced by Sm doping. The detailed growth conditions and characterization can be found in previous publications [34-36].

## Acknowledgements

The development of automated DART optimization workflow (YL, SVK) was supported by the Center for 3D Ferroelectric Microelectronics (3DFeM), an Energy Frontier Research Center funded by the U.S. Department of Energy (DOE), Office of Science, Basic Energy Sciences under Award Number DE-SC0021118.

The work at the University of Maryland was supported by ONR MURI N00014172661, NIST cooperative agreement 70NANB17H301, and DTRA CB11400 MAGNETO, Univ. of Maryland. Growth of PZT samples (SU, STM) was supported by DMR-2025439.

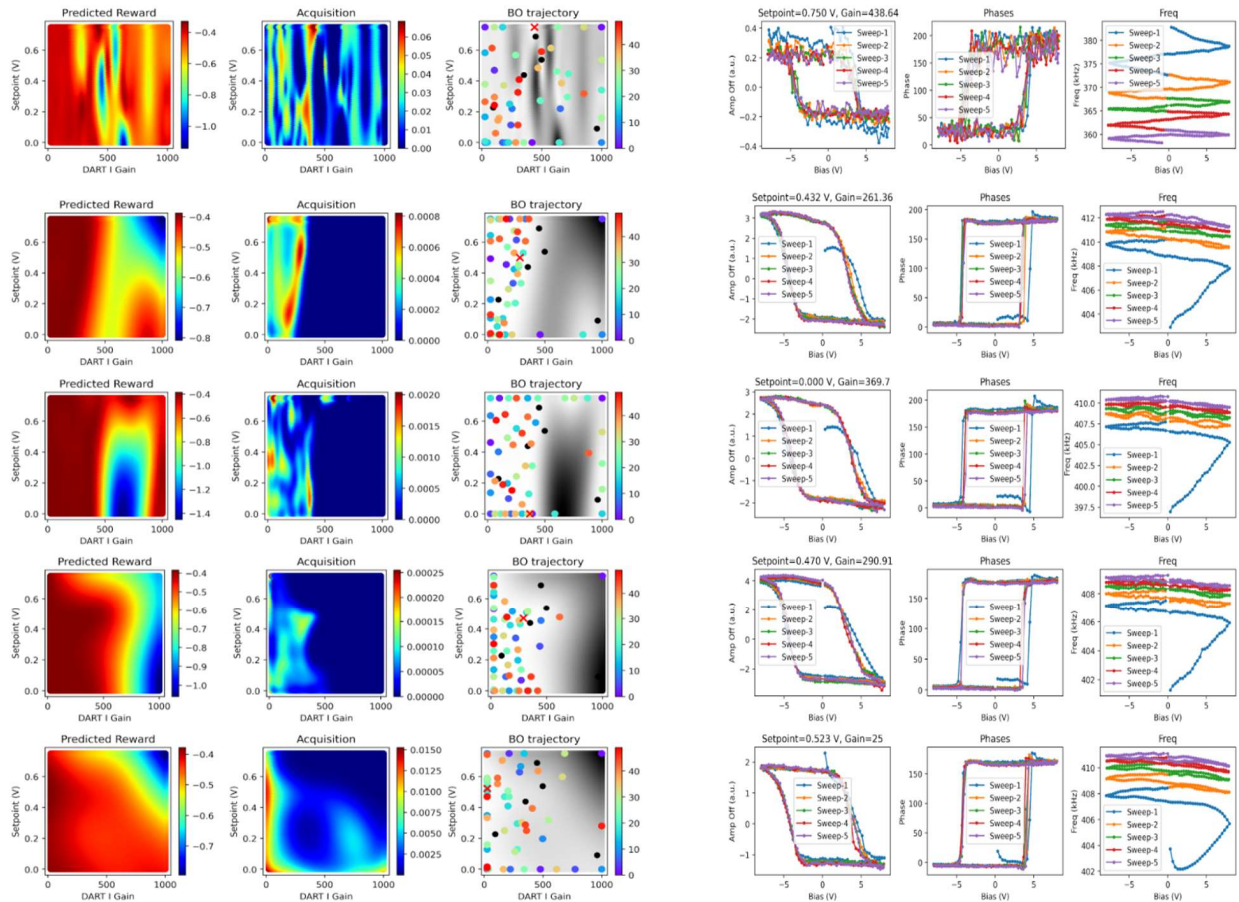

**Figure S1. Consistency of DART optimization with different spectroscopy time measured on PZT.** The optimization process and corresponding optimized DART spectra taken with different spectroscopy times. From top to bottom, the sweep time for a single cycle ranges from 1 s, 1.33 s, 2 s, 2.5 s, to 5 s. Other than the sweeping time, all the other parameters are set identically. Because the number of bias points is kept the same across the five experiments, the data acquisition time at each bias point is inversely proportional to the total sweeping time.

# References

1. García, R. and R. Pérez, *Dynamic atomic force microscopy methods.* Surface Science Reports, 2002. **47**(6): p. 197-301.
2. Bonnell, D.A., et al., *Piezoresponse Force Microscopy: A Window into Electromechanical Behavior at the Nanoscale.* MRS Bulletin, 2009. **34**(9): p. 648-657.
3. Jesse, S., et al., *Direct Mapping of Ionic Transport in a Si Anode on the Nanoscale: Time Domain Electrochemical Strain Spectroscopy Study.* ACS Nano, 2011. **5**(12): p. 9682-9695.
4. Balke, N., et al., *Nanoscale mapping of ion diffusion in a lithium-ion battery cathode.* Nat Nanotechnol, 2010. **5**(10): p. 749-54.
5. Pullar, R.C., et al., *Dielectric measurements on a novel $Ba_{1-x}Ca_xTiO_3$ (BCT) bulk ceramic combinatorial library.* Journal of Electroceramics, 2008. **22**(1-3): p. 245-251.
6. Romanyuk, K., et al., *Single- and Multi-Frequency Detection of Surface Displacements via Scanning Probe Microscopy.* Microscopy and Microanalysis, 2015. **21**(1): p. 154-163.
7. Rodriguez, B.J., et al., *Dual-frequency resonance-tracking atomic force microscopy.* Nanotechnology, 2007. **18**(47): p. 475504.
8. Jesse, S., et al., *The band excitation method in scanning probe microscopy for rapid mapping of energy dissipation on the nanoscale.* Nanotechnology, 2007. **18**(43): p. 435503.
9. Tang, Y.-Y., et al., *A Multiaxial Molecular Ferroelectric with Highest Curie Temperature and Fastest Polarization Switching.* Journal of the American Chemical Society, 2017. **139**(39): p. 13903-13908.
10. Zhang, Y., et al., *Piezoelectric Energy Harvesting Based on Multiaxial Ferroelectrics by Precise Molecular Design.* Matter, 2020. **2**(3): p. 697-710.
11. Li, J., et al., *Composition-induced non-ergodic–ergodic transition and electrocaloric evolution in $Pb_{1-1.5}La\ Zr_{0.8}Ti_{0.2}O_3$ relaxor ferroelectric ceramics.* IET Nanodielectrics, 2019. **2**(4): p. 123-128.
12. Mahmoud, A.E.-r., et al., *Ferroelectric-to-non-ergodic relaxor phase transition of $(Bi_{0.5}Na_{0.3}K_{0.2})TiO_3 – (Ba_{0.8}Ca_{0.2})TiO_3$ lead-free ceramics by $SrTiO_3$ effect.* Journal of Materials Science: Materials in Electronics, 2021. **32**(23): p. 27625-27635.
13. Su, X., et al., *Non-ergodic – ergodic transition and corresponding electrocaloric effect in lead-free bismuth sodium titanate-based relaxor ferroelectrics.* Journal of the European Ceramic Society, 2022. **42**(12): p. 4917-4925.
14. Zhou, C., et al., *Ferroelectric-quasiferroelectric-ergodic relaxor transition and multifunctional electrical properties in $Bi_{0.5}Na_{0.5}TiO_3$-based ceramics.* Journal of the American Ceramic Society, 2018. **101**(4): p. 1554-1565.
15. Starr, M.B. and X. Wang, *Coupling of piezoelectric effect with electrochemical processes.* Nano Energy, 2015. **14**: p. 296-311.
16. Takahashi, Y., et al., *Scanning Probe Microscopy for Nanoscale Electrochemical Imaging.* Analytical Chemistry, 2017. **89**(1): p. 342-357.
17. Tavassol, H., et al., *Electrochemical stiffness in lithium-ion batteries.* Nature Materials, 2016. **15**(11): p. 1182-1187.
18. Yang, S.M., et al., *Mixed electrochemical–ferroelectric states in nanoscale ferroelectrics.* Nature Physics, 2017. **13**(8): p. 812-818.
19. Zhang, G., P. Yu, and S. Shen, *Ferroelectric-like hysteresis loops induced by chemical reaction and flexoelectricity in electrochemical strain microscopy measurements.* Journal of Applied Physics, 2018. **124**(8).

20. Zhang, K., Y. Li, and B. Zheng, *Effects of concentration-dependent elastic modulus on Li-ions diffusion and diffusion-induced stresses in spherical composition-gradient electrodes.* Journal of Applied Physics, 2015. **118**(10).
21. Jones, E.M.C., et al., *Reversible and Irreversible Deformation Mechanisms of Composite Graphite Electrodes in Lithium-Ion Batteries.* Journal of The Electrochemical Society, 2016. **163**(9): p. A1965.
22. Kumar, A., et al., *Nanometer-scale mapping of irreversible electrochemical nucleation processes on solid Li-ion electrolytes.* Scientific Reports, 2013. **3**(1): p. 1621.
23. Luchkin, S.Y., et al., *Reversible and Irreversible Electric Field Induced Morphological and Interfacial Transformations of Hybrid Lead Iodide Perovskites.* ACS Applied Materials & Interfaces, 2017. **9**(39): p. 33478-33483.
24. Bassiri-Gharb, N., S. Trolier-McKinstry, and D. Damjanovic, *Strain-modulated piezoelectric and electrostrictive nonlinearity in ferroelectric thin films without active ferroelastic domain walls.* Journal of Applied Physics, 2011. **110**(12).
25. Bassiri Gharb, N., S. Trolier-McKinstry, and D. Damjanovic, *Piezoelectric nonlinearity in ferroelectric thin films.* Journal of Applied Physics, 2006. **100**(4).
26. Vasudevan, R.K., et al., *Bayesian inference in band excitation scanning probe microscopy for optimal dynamic model selection in imaging.* Journal of Applied Physics, 2020. **128**(5).
27. Proksch, R. and S.V. Kalinin, *Energy dissipation measurements in frequency-modulated scanning probe microscopy.* Nanotechnology, 2010. **21**(45): p. 455705.
28. Liu, Y., et al., *Integration of scanning probe microscope with high-performance computing: Fixed-policy and reward-driven workflows implementation.* Review of Scientific Instruments, 2024. **95**(9).
29. Balandat, M., et al., *BOTORCH: a framework for efficient monte-carlo Bayesian optimization*, in *Proceedings of the 34th International Conference on Neural Information Processing Systems*. 2020, Curran Associates Inc.: Vancouver, BC, Canada. p. Article 1807.
30. Liu, Y., et al., *Exploring Physics of Ferroelectric Domain Walls in Real Time: Deep Learning Enabled Scanning Probe Microscopy.* Adv Sci (Weinh), 2022. **9**(31): p. e2203957.
31. Kalinin, S.V., et al., *Post-Experiment Forensics and Human-in-the-Loop Interventions in Explainable Autonomous Scanning Transmission Electron Microscopy.* Microscopy and Microanalysis, 2023. **29**(Supplement_1): p. 689-690.
32. Liu, Y., et al., *Experimental discovery of structure–property relationships in ferroelectric materials via active learning.* Nature Machine Intelligence, 2022. **4**(4): p. 341-350.
33. Liu, Y., et al., *Exploring the Relationship of Microstructure and Conductivity in Metal Halide Perovskites via Active Learning-Driven Automated Scanning Probe Microscopy.* The Journal of Physical Chemistry Letters, 2023. **14**(13): p. 3352-3359.
34. Nelson, C.T., et al., *Deep learning ferroelectric polarization distributions from STEM data via with and without atom finding.* npj Computational Materials, 2021. **7**(1): p. 149.
35. Nelson, C.T., et al., *Exploring physics of ferroelectric domain walls via Bayesian analysis of atomically resolved STEM data.* Nat Commun, 2020. **11**(1): p. 6361.
36. Ziatdinov, M.A., et al., *Hypothesis Learning in Automated Experiment: Application to Combinatorial Materials Libraries.* Advanced Materials, 2022. **34**(20): p. 2201345.